\documentclass[5p]{elsarticle}

\usepackage{hyperref}
\usepackage[T1]{fontenc}
\usepackage[utf8]{inputenc}
\usepackage{textcomp} 
\usepackage{lmodern} 
\usepackage{multirow}
\usepackage{xcolor}
\usepackage{amsmath}
\usepackage{amsfonts}
\usepackage{breakurl} 
\usepackage[switch]{lineno} 

\usepackage{wasysym}
\usepackage{tabularx}

\graphicspath{{./figures/}}
\biboptions{sort&compress}

\hypersetup{
    breaklinks = true,
    colorlinks = true,
    pdftitle = {}, 
    pdfauthor = {},
    pdfkeywords = {},
    linkcolor = blue,
    citecolor = blue,
    filecolor = black,
    urlcolor = magenta
}

\makeatletter 
\def\@author#1{\g@addto@macro\elsauthors{\normalsize%
    \def\baselinestretch{1}%
    \upshape\authorsep#1\unskip\textsuperscript{%
      \ifx\@fnmark\@empty\else\unskip\sep\@fnmark\let\sep=,\fi
      \ifx\@corref\@empty\else\unskip\sep\@corref\let\sep=,\fi
      }%
    \def\authorsep{\unskip,\space}%
    \global\let\@fnmark\@empty
    \global\let\@corref\@empty  
    \global\let\sep\@empty}%
    \@eadauthor={#1}
}
\makeatother

\newcommand{\lozfept}{L1$\mathrm{_0}$ FePt} 
\newcommand{\feptloz}{FePt L1$\mathrm{_0}$}
\newcommand{\tc}{$T_{\rm C}$}
\newcommand{\opath}{Fe$_{0.5\uparrow}$Fe$_{0.5\downarrow}$Pt}
\newcommand{\tpath}{Fe$_{0.5\uparrow}$Fe$_{0.5\downarrow}$Pt$_{0.5\uparrow}$Pt$_{0.5\downarrow}$}

\newcolumntype{Y}{>{\hsize=.92\hsize\centering\arraybackslash}X}
\newcolumntype{s}{>{\hsize=.08\hsize\centering\arraybackslash}X}
\newcolumntype{b}{>{\hsize=1.08\hsize\centering\arraybackslash}X}

\begin{document}
\begin{sloppypar} 

\title{DFT calculation of intrinsic properties of magnetically hard phase \lozfept}

\author[add1,add2]{Joanna Marciniak\corref{cor1}}
\ead{joanna.marciniak@ifmpan.poznan.pl}
\cortext[cor1]{Corresponding author}

\author[add1,add3]{Wojciech Marciniak}
\author[add1]{Miros\l{}aw Werwi\'nski}

\address[add1]{Institute of Molecular Physics, Polish Academy of Sciences, M. Smoluchowskiego 17, 60-179 Pozna\'n, Poland}

\address[add2]{Institute of Materials Engineering, Poznan University of Technology, Piotrowo 3, 60-965 Pozna\'n, Poland}

\address[add3]{Institute of Physics, Poznan University of Technology, Piotrowo 3, 60-965 Pozna\'n, Poland}

\begin{abstract}
Due to its strong magnetocrystalline anisotropy, \feptloz{} phase is considered as a promising magnetic recording media material. 
Although the magnetic properties of this phase have already been analyzed many times using density functional theory (DFT), we decided to study it again, emphasizing on full potential methods, including spin-polarized relativistic Korringa-Kohn-Rostoker (SPR-KKR) and full-potential local-orbital (FPLO) scheme. 
In addition to the determination of exact values of the magnetocrystalline anisotropy constants \textit{K$\mathrm{_1}$} and \textit{K$\mathrm{_2}$}, the magnetic moments, the Curie temperature, and the magnetostriction coefficient, we focused on the investigation of the magnetocrystalline anisotropy energy (MAE) dependence on the magnetic moment values using the fully relativistic fixed spin moment (FSM) method with various exchange-correlation potentials. 
We present nearly identical MAE($m$) curves near the equilibrium point, along with different equilibrium values of MAE and magnetic moments. 
For a magnetic moment reduced by about 10\%, we determined a theoretical MAE maximum in the ground state (0 K) equal to about 20.3 MJ\,m$^{-3}$ and independent of the choice of the exchange-correlation potential form.
These calculations allow us to understand the discrepancies between the previous MAE results for different exchange-correlation potentials.

\end{abstract}


\maketitle

\section{Introduction}

In 2011, the interest in rare-earth-free hard magnetic materials increased significantly due to the dramatic rise in prices of rare earth elements known as the rare earth crisis~\cite{bourzacRareearthCrisis2011}. 
Several articles summarize results of research on magnetic materials free of rare earths, among others, see Refs.~\cite{kuzminHighperformancePermanentMagnets2014} and \cite{skomskiMagneticAnisotropyHow2016}. 
One of the promising materials is the ordered \lozfept{} alloy, for which measurements at 4.2~K revealed an unusually high value of magnetocrystalline anisotropy constant $K_1$ of 1.92~meV\,f.u.$\mathrm{^{-1}}$ (11.0~MJ\,m$\mathrm{^{-3}}$)~\cite{haiMagneticPropertiesHard2003}. 
Another interesting property of the \feptloz{} phase, concerning permanent-magnet applications, is the relatively high Curie temperature (\tc{}) of about 775~K estimated for the bulk material from experiments on nanoparticles~\cite{rongSizeDependentChemicalMagnetic2006}. 
In the literature, the crystallographic parameters of the tetragonal \feptloz{} structure vary: $a$~=~3.85--3.88~\AA, $c$~=~3.74--3.79~\AA, and the ratio $c/a$~=~0.966--0.981~\cite{ravindranLargeMagnetocrystallineAnisotropy2001, luFirstprinciplesStudyMagnetic2010, klemmerStructuralStudiesL102002, alsaadStructuralElectronicMagnetic2020, luoEffectStoichiometryMagnetocrystalline2014}. 
Values of the total spin magnetic moments in the range 3.06--3.24~$\mathrm{\mu_B}$ per formula unit were obtained to date~\cite{ayazkhanMagnetocrystallineAnisotropyFePt2016, wollochInfluenceAntisiteDefects2017, burkertMagneticAnisotropyL102005, alsaadStructuralElectronicMagnetic2020}.
The magnetostriction coefficient $\mathrm{\lambda_{001}}$ varies between approximately 3.4$\mathrm{\times 10^{-5}}$~\cite{spadaXrayDiffractionMossbauer2003} and 6.4$\mathrm{\times}$10$\mathrm{^{-5}}$~\cite{aboafMagneticTransportStructural1984} for disordered films of FePt alloys.
For comparison, the highest value of magnetostriction coefficient, equal to 1.1--1.4$\mathrm{\times 10^{-3}}$, was recorded for the Terfenol-D alloy~\cite{sandlundMagnetostrictionElasticModuli1994}.

\begin{figure}[t]
\centering
\includegraphics[clip,width=0.75\columnwidth]{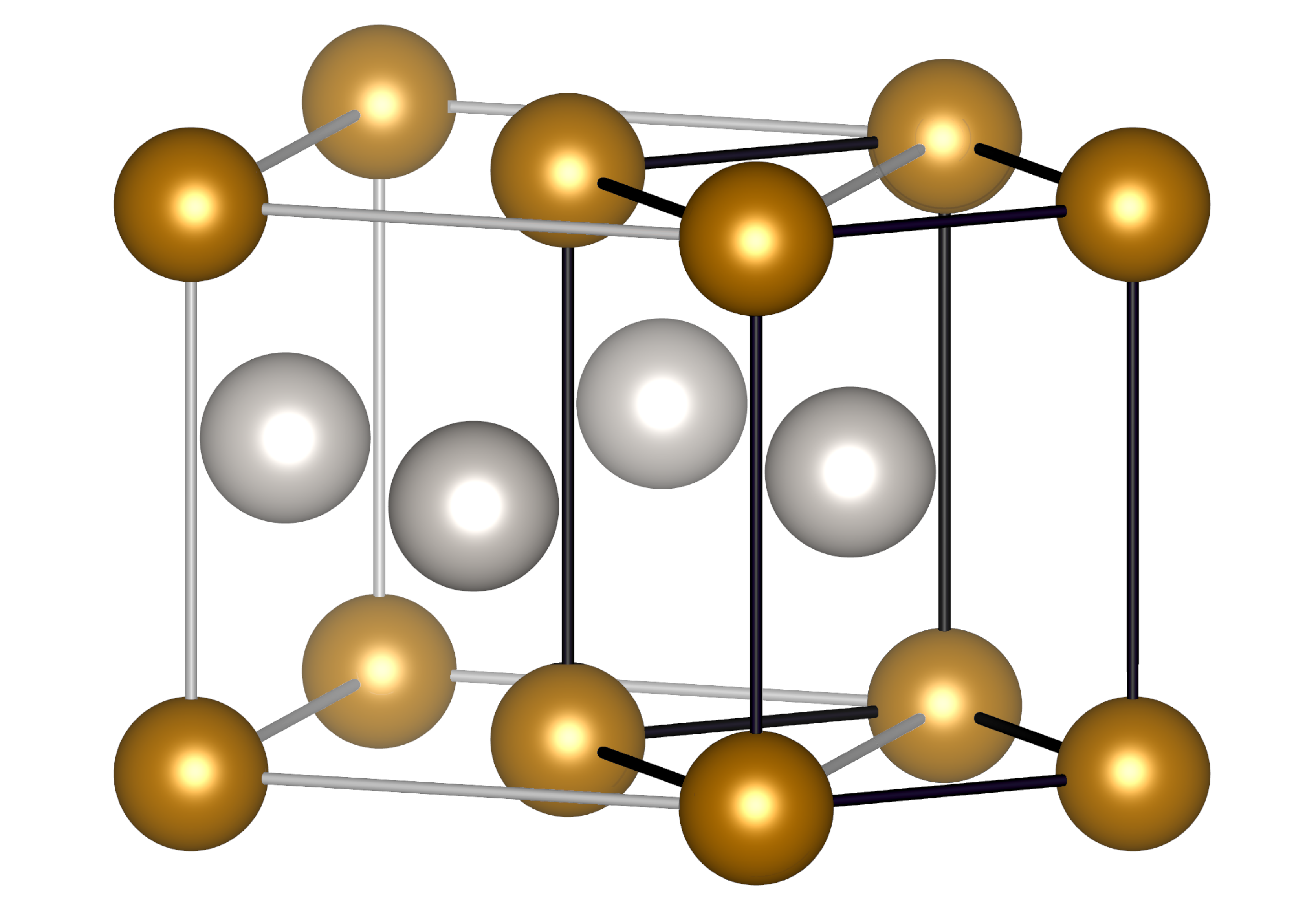}
\caption{\label{fig-l10_structure}
	Crystal structure of \feptloz{} phase, in face-centered tetragonal (gray lines) and body-centered tetragonal (black lines) representations. The latter was used for calculations in this work. 
{\sc Vesta} code was used for visualization~\cite{mommaVESTAThreedimensionalVisualization2008a}.
}
\end{figure}

Our aim in the following work is twofold. 
Firstly, we want to present a thorough benchmark study regarding the basic magnetic properties -- magnetocrystalline anisotropy energy (MAE), magnetocrystalline anisotropy constants $K_1$ and $K_2$, magnetic moments~\cite{ayazkhanMagnetocrystallineAnisotropyFePt2016, wollochInfluenceAntisiteDefects2017} and Curie temperature \cite{rongCurieTemperaturesAnnealed2007} -- of this interesting, already well studied phase. 
To obtain these parameters, we utilize two density functional theory (DFT) codes and a possibly broad spectrum of exchange-correlation functionals (described in Sec.~\ref{calc}). 
Secondly and more importantly, we try to broaden the understanding of the magnetic properties of the presented alloy by fixed spin moment analysis and magnetostriction coefficient calculations, though the comparative experimental results of the latter present in literature are sparse~\cite{spadaXrayDiffractionMossbauer2003}, especially for the bulk alloy \cite{aboafMagneticTransportStructural1984}.

\section{Calculations' details}
\label{calc}

In this work, \textit{ab initio} calculations were performed using FPLO18, FPLO5 (full-potential local-orbital)~\cite{eschrigChapter12Relativistic2004, koepernikFullpotentialNonorthogonalLocalorbital1999}, and SPR-KKR 7.7.1 (spin-polarized relativistic Korringa-Kohn-Rostoker) codes~\cite{ebertMunichSPRKKRPackage2017, ebertCalculatingCondensedMatter2011}.

The FPLO18 code was used for calculations necessary to determine spin and orbital magnetic moments, MAE, magnetocrystalline anisotropy constants \textit{K$\mathrm{_1}$} and \textit{K$\mathrm{_2}$}, and the magnetostriction coefficient. 
Dependence of MAE on the spin magnetic moment was performed using the fixed spin moment (FSM) method.

The FPLO5 code was used as a base for the material's Curie temperature calculations, and the SPR-KKR code was used to obtain \tc{}, MAE and magnetocrystalline anisotropy constants for cross-reference. 
The main reason for the use of FPLO5 and SPR-KKR, compared to FPLO18, is that both codes have the chemical disorder implemented by means of the coherent potential approximation (CPA)~\cite{sovenCoherentPotentialModelSubstitutional1967}, which allows utilizing the approach described further to obtain Curie temperature with relatively low computational effort.

The high precision of FPLO is due, among other factors, to the implementation of a full-potential approach that does not incorporate shape approximation into the crystalline potential and to the expansion of extended states in terms of localized atomic-like basis orbitals~\cite{koepernikFullpotentialNonorthogonalLocalorbital1999, eschrigEssentialsDensityFunctional2004}. 
The use of full potential is particularly essential for the accurate determination of such a subtle quantity as MAE, and the MAE results obtained for FePt using this method are considered among the most accurate~\cite{wollochInfluenceAntisiteDefects2017}. 
SPR-KKR 7.7.1 also has the ability for full potential calculation, which, oppositely to FPLO, does not lie in the basic principles of the code.

All calculations were performed after a proper optimization of the system geometry utilizing Perdew-Burke-Ernzerhof (PBE) exchange-correlation functional~\cite{perdewGeneralizedGradientApproximation1996} in corresponding codes.
The calculations were made for the body-centered representation of the \feptloz{} structure, see Fig.~\ref{fig-l10_structure}, belonging to the \textit{P}4\textit{/mmm} space group, with Fe atoms occupying the (0, 0, 0) position and Pt atoms occupying the (0.5, 0.5, 0.5) site. 
Lattice parameters of $a=3.88/\sqrt{2}$~\AA{} and $c=3.73$~\AA{}, derived from face-centered tetragonal (fct) structure of Alsaad \textit{et al.}~\cite{alsaadStructuralElectronicMagnetic2020}, were used as initial values for geometry optimization. 
Final lattice parameters used in further calculations are: $a=2.74\approx3.87/\sqrt{2}$~\AA, $c=3.76$~\AA{} and $a=2.74\approx3.87/\sqrt{2}$~\AA, $c=3.74$~\AA{}, optimized with FPLO and SPR-KKR codes, respectively, see Fig.~\ref{fig-optimisation}.

Main calculations were performed using the Perdew-Burke-Ernzerhof (PBE) (FPLO18 and SPR-KKR), Vosko-Wilk-Nusair (VWN) (SPR-KKR)~\cite{voskoAccurateSpindependentElectron1980} and Perdew-Wang 92 (PW92) (FPLO18 and FPLO5) \cite{perdewAccurateSimpleAnalytic1992} exchange-correlation potentials. 
Additional results of Curie temperature and MAE dependence on the magnetic moment were obtained in FPLO5 and FPLO18, respectively. 
In both codes, we used additionally von Barth-Hedin (vBH)~\cite{barthLocalExchangecorrelationPotential1972} and Perdew-Zunger (PZ) \cite{perdewSelfinteractionCorrectionDensityfunctional1981} potentials together with the exchange only approximation.

All calculations in FPLO18 were performed in the fully relativistic approach for a $56\times 56\times 42$ \textbf{k}-point mesh with energy convergence criterion at the level of 10$^{-8}$~Ha and electron density convergence criterion of $10^{-6}$, which yields an accurate value of the system's MAE within a reasonable computation time. 
Curie temperature calculations were performed in the scalar-relativistic approach in CPA for a $12\times 12\times 12$ \textbf{k}-point mesh and the same energy and density convergence criteria.

The convergence criterion at the level of 10$^{-6}$ and 5~000 \textbf{k}-points per reduced Brillouin zone, around 75~000 \textbf{k}-points in a full Brillouin zone, were used to perform fully relativistic calculations in SPR-KKR. 
All calculations were carried out in the full-potential (FP) approach with the angular momentum cut-off criterion set at the level of \textit{l$\mathrm{_{max}}$}~=~5 (parameter NL~=~6 in the SPR-KKR configuration file) unless otherwise stated. 
The atomic sphere approximation (ASA) calculations in SPR-KKR were performed for comparison, and all computational parameters were consistent between both approaches.

The Curie temperature of the system was determined based on the disordered local moment (DLM) theory~\cite{stauntonDisorderedLocalMoment1984} and assuming that the energy difference between the ferromagnetic state (ordered spin magnetic moments) and the paramagnetic state (disordered spin magnetic moments -- modelled based on two types of CPA atomic sites with antiparallel magnetic moments) is proportional to the thermal energy needed for the ferromagnetic--paramagnetic transition. 
In order to determine the Curie temperature, we use the formula~\cite{gyorffyFirstprinciplesTheoryFerromagnetic1985, bergqvistTheoreticalStudyHalfmetallic2007a}:

\begin{equation}\label{eq:dlm}
k_\mathrm{B}T_\mathrm{C} = \frac{2}{3} \frac{E_\mathrm{DLM} - E_\mathrm{FM}}{n},
\end{equation}

\noindent where $E_\mathrm{DLM}$ is the total energy of the DLM configuration, $E_\mathrm{FM}$ is the total energy of the ferromagnetic configuration, $n$ is the total number of magnetic atoms, and $k_\mathrm{B}$ is the Boltzmann constant.

MAE was determined by the formula:

\begin{equation}\label{eq:MAE}
\mathrm{MAE} = E(\theta=90^\circ) - E(\theta=0^\circ),
\end{equation}

\noindent where $\theta$ is the angle between the magnetization direction and the c axis. 
We determined the hard magnetic axis in the computational cell to be [1 1 0] ([1 0 0] in the standard face-centered representation). 
However, difference between energy values in the $x$\nobreakdash-$y$~plane is insignificant in our case and omitted in further study. 
In order to determine the magnetocrystalline anisotropy constants $K_1$ and $K_2$, the dependence MAE($\theta$) was used, which for a tetragonal cell takes the approximated form:

\begin{equation}\label{eq:k1-k2}
\mathrm{MAE} = K_1 \sin^{2}\theta + K_2 \sin^4\theta.
\end{equation}

\begin{figure}[t]
\centering
\includegraphics[clip,width=1.0\columnwidth]{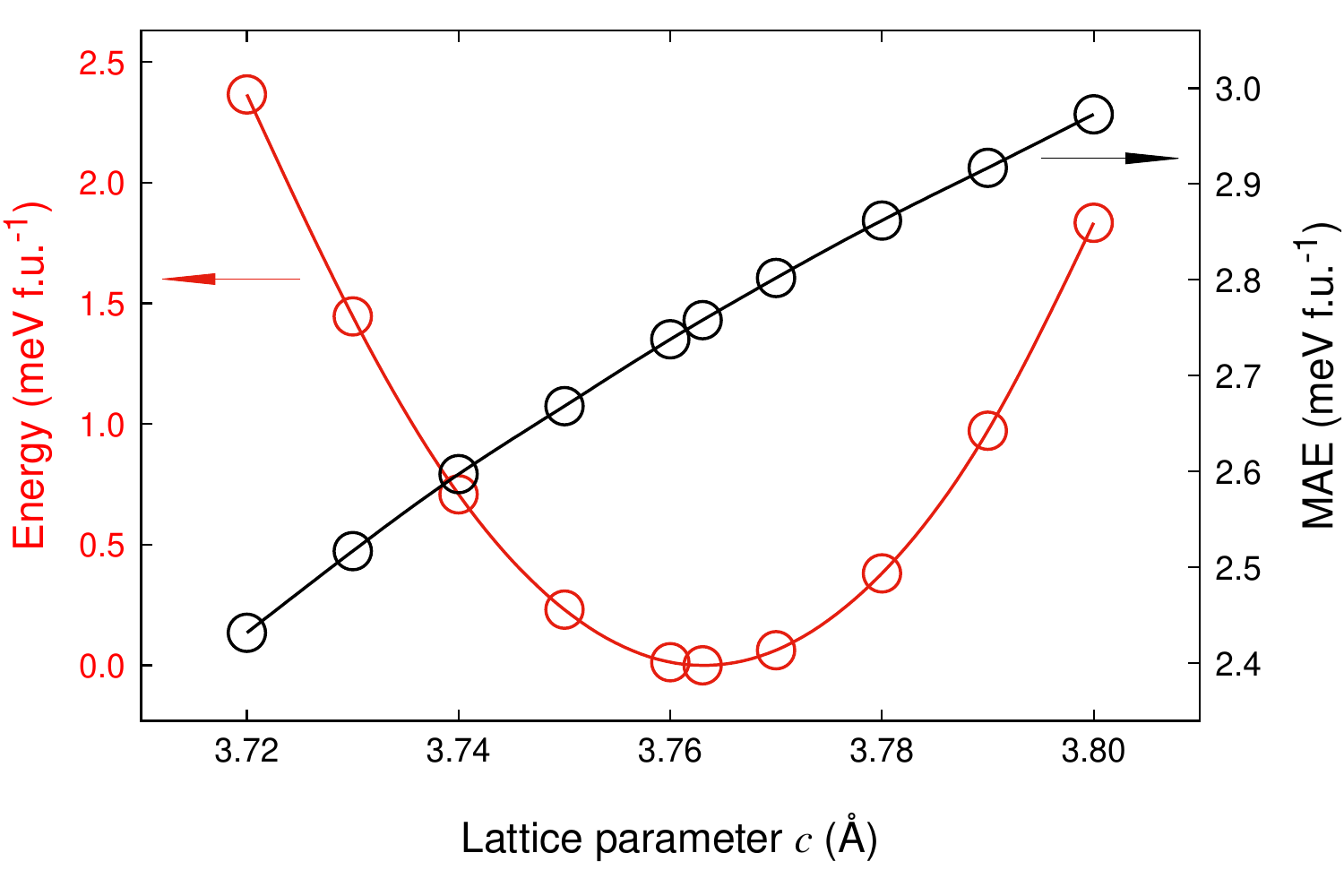}
\caption{\label{fig-magnetostriction-coeffitient}
Dependence of the calculated total energy and magnetocrystalline anisotropy energy on the lattice parameter of \lozfept{} \textit{c} calculated with FPLO18 code.
}
\end{figure}

In order to calculate the magnetostriction coefficient $\lambda_{001}$, MAE and energy dependence on the lattice parameter $c$ were fit with linear and quadratic functions, respectively~\cite{wuSpinOrbitInduced1999, werwinskiInitioStudyMagnetocrystalline2017}, as presented in Fig.~\ref{fig-magnetostriction-coeffitient}.
The relation between energy and MAE \textit{versus} magnetization angle $\theta$ can be presented as follows:

\begin{equation}\label{eq:b}
\begin{aligned}
E(\theta = 0^{\circ})&= \alpha c^2+\beta c+\gamma; \\
E(\theta = 90^{\circ})&= \alpha c^2+\beta c+\gamma+\mathrm{MAE}(c).
\end{aligned}
\end{equation}

Considering the lattice parameter elongation derivative of the MAE for the optimized lattice parameter ($c=3.76$~\AA), the magnetostriction coefficient $\mathrm{\lambda_{001}} $ takes the form:

\begin{equation}\label{eq:magnetostriction}
	\lambda_{001} = - \frac{2}{3} \frac{\frac{\mathrm{d}(\mathrm{MAE})}{\mathrm{d}{c}}}{\beta}.
\end{equation}

\section{Results and Discussion}

\begin{figure}[t]
\centering
\includegraphics[clip,width=1\columnwidth]{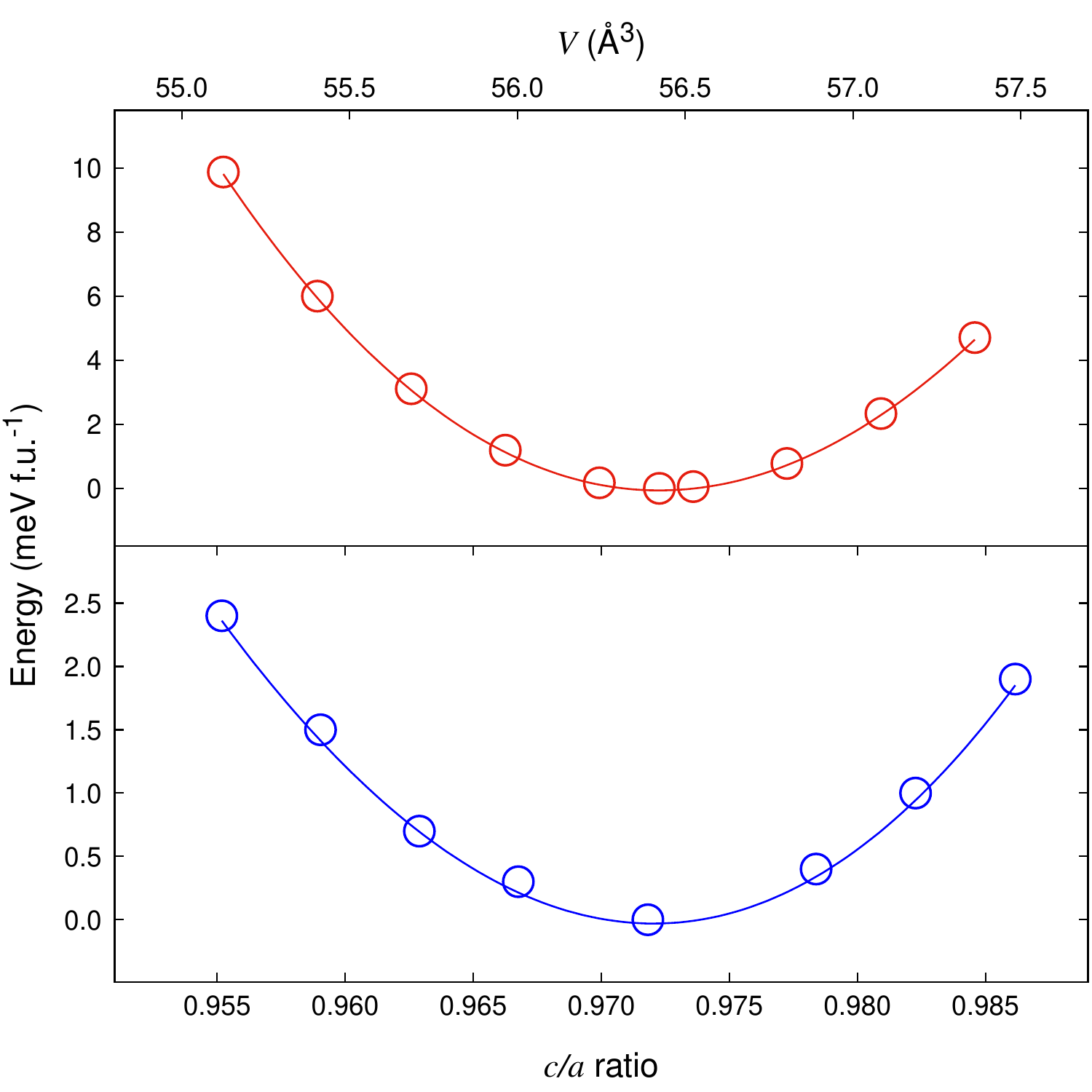}
\caption{\label{fig-optimisation}
Optimization of the \feptloz{} phase geometry in FPLO18 code using the PBE potential. 
Energy minimum shifted to the lowest obtained energy. 
The top figure (red line) depicts the energy \textit{versus} volume dependence for fixed $c/a$ ratio equal to 0.972, and the bottom figure (blue line) presents the energy \textit{versus} $c/a$ ratio for constant volume equal to 56.42~\AA$^3$ (for fct cell). 
Obtained lattice parameters are \textit{a}~=~3.87~\AA{} and \textit{c}~=~3.76~\AA, see the fct structure presented in Fig.~\ref{fig-l10_structure}.
}
\end{figure}

{\renewcommand{\arraystretch}{1.2}
\begin{table*}[t]
\centering
\caption{\label{tab-results}
Spin magnetic moments $m_s$, orbital magnetic moments $m_l$, and total magnetic moments $m$ (in $\mu_B$), as well as magnetocrystalline anisotropy energy MAE (in MJ\,m$\mathrm{^{-3}}$, and meV\,f.u.$^{-1}$) and magnetocrystalline anisotropy constants $K_1$ and $K_2$ (in MJ\,m$\mathrm{^{-3}}$) of the \lozfept{}. 
The values of MAE, $K_1$ and $K_2$ in MJ\,m$\mathrm{^{-3}}$ from~\cite{ayazkhanMagnetocrystallineAnisotropyFePt2016} were recalculated by authors. 
SPR-KKR $l_{\rm max}$ cutoff was set to 5, unless otherwise stated.
}
\centering
\begin{tabular}{ccccccccccc}
\hline \hline
                          & $m_{s,Fe}$ & $m_{l,Fe}$ & $m_{s,Pt}$ & $m_{l,Pt}$ & $m$ & $K_1$ & $K_2$ & MAE (MJ\,m$\mathrm{^{-3}}$) & MAE (meV\,f.u.$^{-1}$) \\
\hline
FPLO PBE		  & 2.95 & 0.065 & 0.22 & 0.057 & 3.28 & 15.3 & 0.39 & 15.66 & 2.76 \\
FPLO PW 92		  & 2.85 & 0.067 & 0.24 & 0.057 & 3.21 & -- & -- & 17.01 & 2.99 \\
FPLO PZ		 	  & 2.81 & 0.067 & 0.24 & 0.057 & 3.17 & -- & -- & 17.71 & 3.12 \\
FPLO vBH 		  & 2.81 & 0.067 & 0.24 & 0.057 & 3.17 & -- & -- & 17.69 & 3.12 \\
FPLO Exchange only	  & 3.07 & 0.066 & 0.22 & 0.056 & 3.41 & -- & -- & 13.03 & 2.29 \\
SPR-KKR PBE ASA           & 3.07 & 0.072 & 0.33 & 0.046 & 3.51 & 16.4 & 0.77 & 17.14 & 2.99 \\
SPR-KKR PBE FP            & 2.98 & 0.064 & 0.33 & 0.047 & 3.42 & 16.7 & 0.47 & 17.17 & 3.00 \\
SPR-KKR VWN ASA           & 2.95 & 0.071 & 0.33 & 0.041 & 3.39 & 21.4 & -0.13 & 21.29 & 3.72 \\
SPR-KKR VWN FP            & 2.89 & 0.066 & 0.34 & 0.046 & 3.33 & 17.9 & 0.76 & 18.63 & 3.25 \\
SPR-KKR VWN & \multirow{2}{*}{2.91} & \multirow{2}{*}{0.073} & \multirow{2}{*}{0.32} & \multirow{2}{*}{0.043} & \multirow{2}{*}{3.35} & \multirow{2}{*}{17.6} & \multirow{2}{*}{0.29} & \multirow{2}{*}{17.98} & \multirow{2}{*}{3.14} \\
ASA $l_{max}=3$		  & & & & & & & & & \\
SPR-KKR VWN & \multirow{2}{*}{2.83} & \multirow{2}{*}{0.065} & \multirow{2}{*}{0.34} & \multirow{2}{*}{0.044} & \multirow{2}{*}{3.28} & \multirow{2}{*}{17.5} & \multirow{2}{*}{0.54} & \multirow{2}{*}{17.96} & \multirow{2}{*}{3.01} \\
ASA $l_{max}=3$~\cite{ayazkhanMagnetocrystallineAnisotropyFePt2016} & & & & & & & & & \\
VASP PBE~\cite{wollochInfluenceAntisiteDefects2017}	& 2.83 & 0.056 & 0.39 & 0.044 & 3.32 & 15.3 & 0.74 & 15.7 & 2.74 \\
\hline \hline
\end{tabular}
\end{table*}
}

\begin{figure}[t]
\centering
\includegraphics[clip,width=1.0\columnwidth]{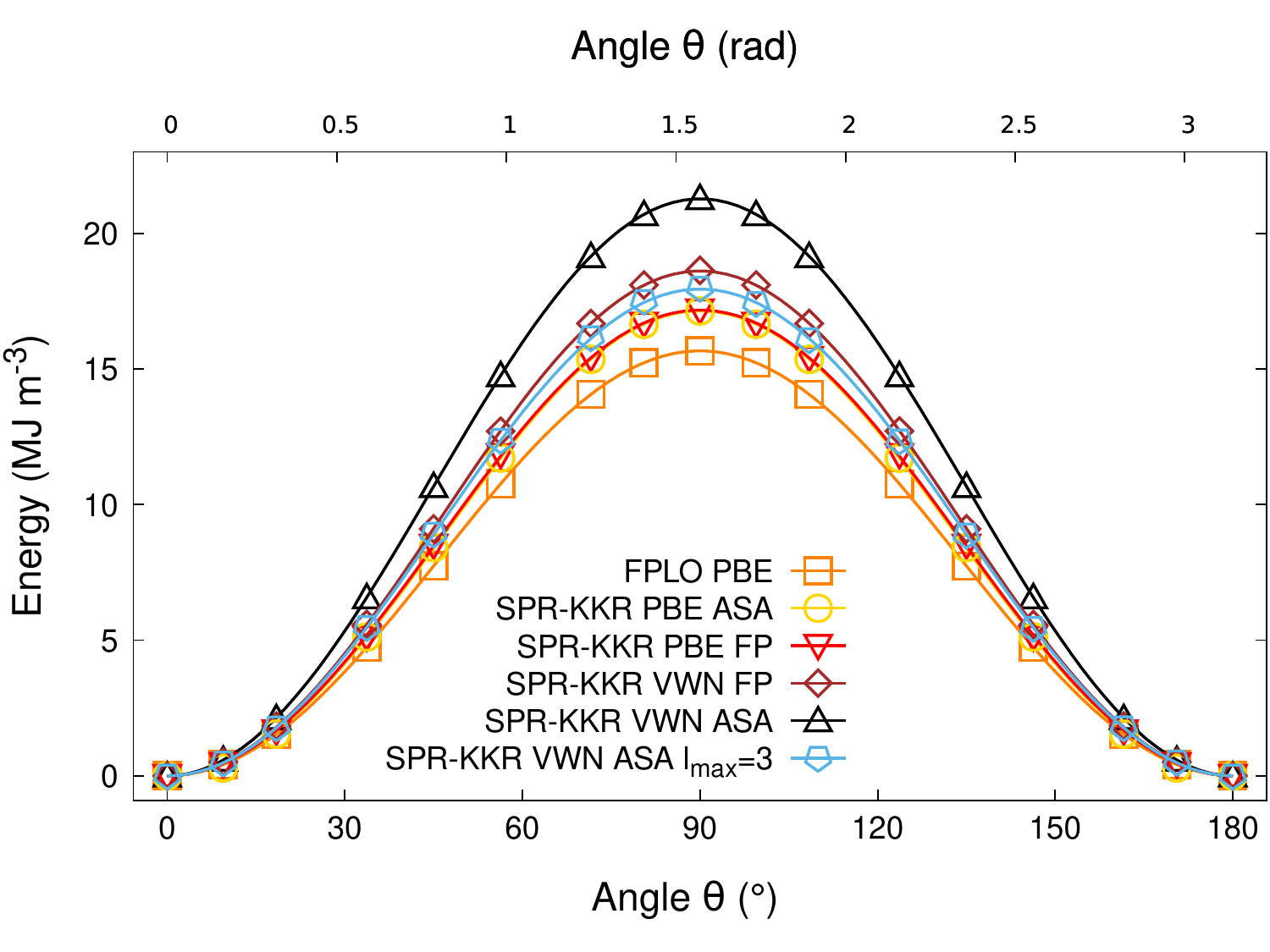}
\caption{\label{fig-coeff-k1-k2}
Energy of the \lozfept{} phase, as a function of the angle between magnetization direction and the c axis, calculated with various exchange-correlation potentials in FPLO18 and SPR-KKR 7.7.1. 
Exact values of magnetocrystalline anisotropy constants $K_1$ and $K_2$ are presented in Tablee~\ref{tab-results}. 
SPR-KKR $l_{\rm max}$ cutoff was set to 5, unless stated otherwise.
}
\end{figure}

Before the actual calculations, the geometry of the crystal structure was optimized. 
The procedure consisted of a simple least-squares third-order function fit to the $E(V)$ dependency under a constant $c/a$ ratio and then the same kind of fit to the energy dependence $E(c/a)$ at a constant volume corresponding to the minimum energy devised before. 
For the FPLO18 with PBE, the equilibrium volume of the fct cell $V_{eq}$ is equal to $56.42$\,\AA $\mathrm{^3}$ and the equilibrium $c/a$ ratio is equal to $0.972$, see Fig.~\ref{fig-optimisation}.
For the SPR-KKR code and PBE exchange-correlation functional, $a$ value is lower by 0.5\permil, whereas $c$ is smaller by 7.2\permil{} than that resultant from FPLO calculations.
All results provide very good agreement with previous experimental and calculations results \cite{ravindranLargeMagnetocrystallineAnisotropy2001, luFirstprinciplesStudyMagnetic2010, klemmerStructuralStudiesL102002, alsaadStructuralElectronicMagnetic2020, luoEffectStoichiometryMagnetocrystalline2014}.

For FPLO18-PBE, the spin magnetic moment $m_s=\mathrm{2.95 \mu_B}$ and the orbital magnetic moment $m_l = \mathrm{0.065 \mu_B} $ were observed on Fe atoms. 
On Pt atoms, these moments were $m_s=\mathrm{0.22 \mu_B}$ and $m_l=\mathrm{0.057\mu_B}$. 
For values obtained using methods other than the FPLO18 code with PBE, see Table~\ref{tab-results}. 
The sum of magnetic moments in the cell varies between 3.17 and 3.512$\mu_B$, a discrepancy of about 10\% relative to either of the values.
The calculated magnetic moments, presented also in the Table~\ref{tab-results}, are consistent with the values published previously~\cite{ayazkhanMagnetocrystallineAnisotropyFePt2016, wollochInfluenceAntisiteDefects2017, burkertMagneticAnisotropyL102005, alsaadStructuralElectronicMagnetic2020}.

As mentioned in Sec.~\ref{calc}, Curie temperatures were calculated with the DLM method using Eq.~\ref{eq:dlm} for all exchange-correlation potentials implemented in FPLO5 and for VWN and PBE exchange-correlation potentials in SPR-KKR. 
Two approaches were used to simulate the paramagnetic phase. 
In the first one, we assume no Pt contribution to the total magnetic moment and introduce coexisting antiparallel spin orientations on the Fe atom only, i.e. \opath. 
In the second approach, we introduce coexisting antiparallel spin orientations on both Fe and Pt atoms, i.e. \tpath. 
\tc{} values obtained using both approaches are mostly identical, as in both cases, the Pt atoms are going to be in a non-magnetic state (magnetic moments equal to or near 0$\mu_{\rm B}$), as expected.
{All Pt moments are presented in Table~\ref{tab-temperatures}.}

Results of calculated Curie temperatures vary considerably for different exchange-correlation potentials, from about 430~K to 900~K. 
All of them are also presented in Table~\ref{tab-temperatures}. 
Overall, the magnetic moments on Pt in the DLM {stable} states are minor, if any, and has a negligible physical impact on the \tc{} value.
{As shown in Table~\ref{tab-temperatures}, in all considered cases the magnetic moment on Pt atom is below numerical accuracy.}
\tc{}'s obtained with the VWN potential are low, suggesting that this form of exchange-correlation potential in the local density approximation is not adequate to properly capture interatomic exchange interactions and thus \tc{} values. 
On the other hand, FPLO exchange only approach seems to be sufficient. 
We were unable to obtain a stable DLM state using SPR-KKR with PBE exchange-correlation potential, so \tc{} for this case was not included in Table~\ref{tab-temperatures}. 
The results, except for the one obtained in SPR-KKR VWN, agree well with the estimation from experiment by Rong \textit{et al.} (775~K)~\cite{rongCurieTemperaturesAnnealed2007}. 
However, the DLM approach is known to overestimate values of \tc{} considerably~\cite{gyorffyFirstprinciplesTheoryFerromagnetic1985, ebertCalculatingCondensedMatter2011}, so slightly higher values could be expected.

{\renewcommand{\arraystretch}{1.7}
\begin{table}[h]
\centering
\caption{\label{tab-temperatures}
Curie temperatures and total magnetic moments on Pt atoms in L1$\mathrm{_0}$ {\tpath{}} obtained using SPR-KKR and FPLO5 codes.}
\centering
{
\begin{tabularx}{\columnwidth}{
		p{\dimexpr.52\linewidth-2\tabcolsep-1.333\arrayrulewidth}
		p{\dimexpr.24\linewidth-2\tabcolsep-1.333\arrayrulewidth}
		p{\dimexpr.24\linewidth-2\tabcolsep-1.333\arrayrulewidth}
	}
\hline \hline
			& \tc{} (K)	& $m_{Pt}$ ($\mu_\mathrm{B}$)	\\
\hline
 SPR-KKR VWN ASA	& 426		& 0				\\
 SPR-KKR VWN		& 595		& 0				\\
\hline
 FPLO5 Exchange only	& 878		& $< 10^{-6}$			\\
 FPLO5 PW92		& 796		& $< 10^{-6}$			\\
 FPLO5 PZ		& 754		& $< 10^{-6}$			\\
 FPLO5 vBH		& 756		& 0				\\
\hline \hline
\end{tabularx}
}
\end{table}}

The magnetocrystalline anisotropy constants $K_1$ and $ K_2 $ are summarized in Table~\ref{tab-results}, and the results obtained are also shown graphically in Fig.~\ref{fig-coeff-k1-k2}, along the fits of Eq.~\ref{eq:k1-k2} for each data series. 
These data series were performed for the angle between the easy magnetization axis and defined magnetization direction in the $\langle 0; \pi \rangle$ range.
Dependencies are symmetrical in relation to $\theta = \frac{\pi}{2}$, as expected.
Eq.~\ref{eq:k1-k2} was fitted to the whole angular range to improve the accuracy of the fit.
Presented values do not differ significantly, except for the calculations with the SPR-KKR code using the VWN potential in ASA. 
This difference means that the calculations using this potential depend relatively strong on the applied approach -- ASA or FP, as mentioned earlier.
Out of other results, these from FPLO18 and FP PBE SPR-KKR are the lowest, indicating better agreement to experiment.
The difference between FPLO and SPR-KKR can be ascribed to the implemented basis.
Moreover, our convergence tests suggested that the minimal useable angular momentum expansion in the system is \textit{l$\mathrm{_{max}}$}~=~4 (parameter NL~=~5 in the SPR-KKR configuration file), and $l_{max}=5$, used in our work, gave yet further noticeable improvement. 
Taking into account the above, we are convinced that LDA-ASA approach may be insufficient to capture the magnetism of \lozfept{} properly, which is exemplified by our result for SPR-KKR VWN approach in a higher basis of $l_{max}=5$.

All obtained magnetocrystalline anisotropy constants agree with earlier results of Khan \textit{et al.} [$K_1 = 17.2$~MJ\,m$\mathrm{^{-3}}$ (3.01 meV\,f.u.$^{-1}$), $K_2 = 0.53$~MJ\,m$\mathrm{^{-3}}$ (0.09 meV\,f.u.$^{-1}$)] and Wolloch \textit{et al.} [$K_1 = 15.3$~MJ\,m$\mathrm{^{-3}}$ (2.67 meV\,f.u.$^{-1}$), $K_2 = 0.74$~MJ\,m$\mathrm{^{-3}}$ (0.13 meV\,f.u.$^{-1}$)]~\cite{ayazkhanMagnetocrystallineAnisotropyFePt2016, wollochInfluenceAntisiteDefects2017}. 
However, the obtained magnetocrystalline anisotropy constants with $K_1$ values in range 15.3--21.4 MJ\,m$^{-3}$ differ significantly from the values obtained experimentally ($K_1 = 11$~MJ\,m$\mathrm{^{-3}}$)~\cite{haiMagneticPropertiesHard2003}.
Origin of the difference can be connected with imperfections of both, calculations and experiments. 
Weaknesses of calculations are the impossibility of accurately determination of the exchange-correlation potential and difficulty in the exact evaluation of spin-orbit interactions~\cite{daalderopFirstprinciplesCalculationMagnetocrystalline1990}. 
On the other hand, experiments were performed on nanoparticles or bulk samples with inhomogeneous microstructures that impact obtained properties~\cite{liuMagneticPropertiesL12018}. 
Thin films are deposited on substrates, which can induce strain in the examined structure and introduce some differences in properties~\cite{daalderopFirstprinciplesCalculationMagnetocrystalline1990}. 
The most important fact is that the ground state is not measured in experiments~\cite{strangeFirstPrinciplesTheory1991}, so these values cannot be directly compared.
It was checked by Daalderop \textit{et al.}~\cite{daalderopFirstprinciplesCalculationMagnetocrystalline1990} and Strange \textit{et al.}~\cite{strangeFirstPrinciplesTheory1991} that all the approximations in the intrinsic properties calculations can not account for the discrepancy.

\begin{figure}[t]
\centering
\includegraphics[clip,width=1.0\columnwidth]{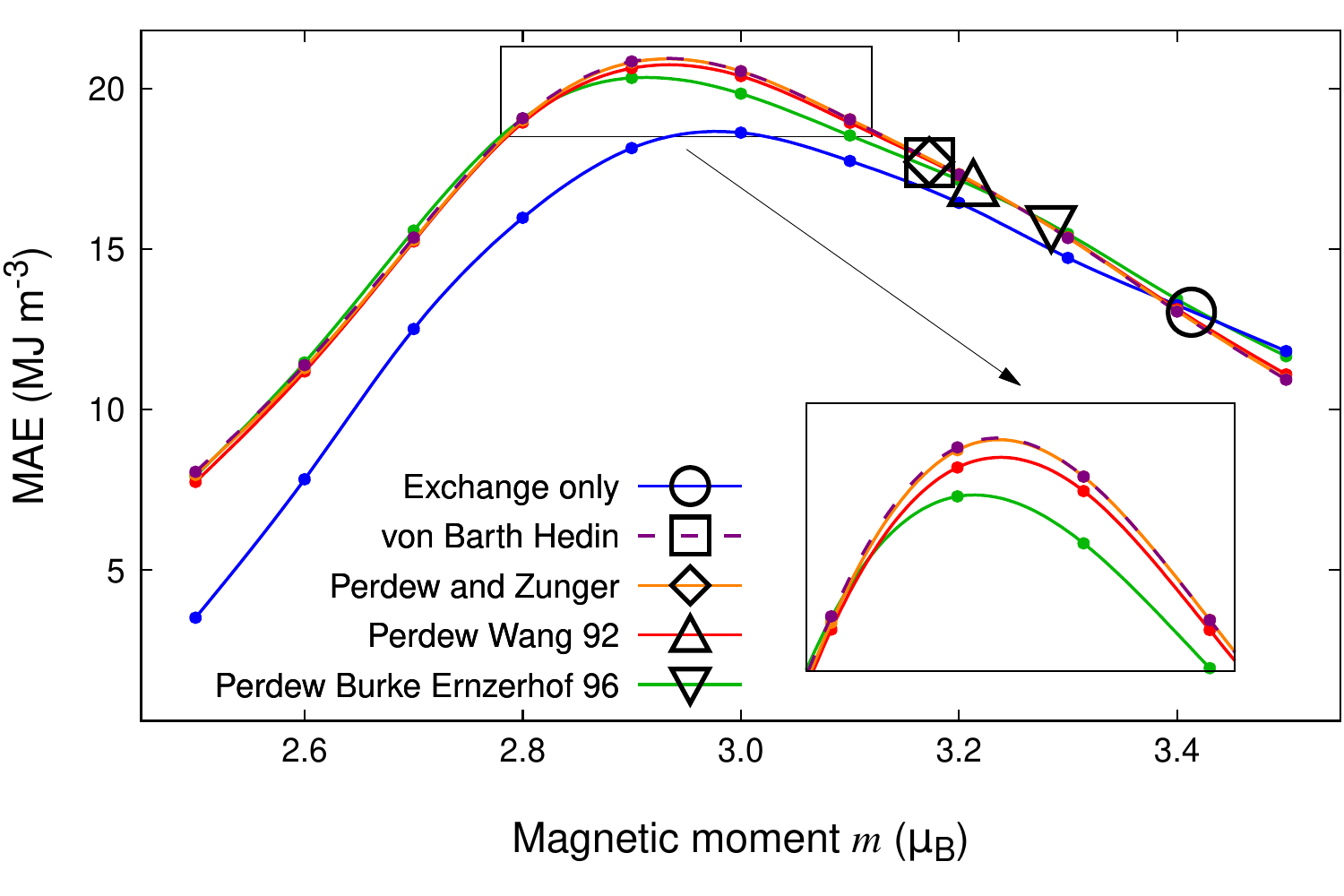}
\caption{\label{fig-fixed-spin}
Fixed spin moment MAE dependencies (lines) resultant from the calculations with different exchange-correlation potentials performed in FPLO18. Equilibrium magnetocrystalline anisotropy values \textit{versus} equilibrium total magnetic moments are marked with empty symbols.
}
\end{figure}

The combination of the FSM method with fully relativistic calculations further allows testing the dependence of the MAE on the total magnetic moment, which allows for determining a hypothetical maximum MAE depending on the magnetic moment for a given material. 
The MAE values obtained with various exchange-correlation potentials in FPLO18 code are 13.03--17.71 MJ\,m$^{-3}$ and correlate with the total magnetic moment in the range 3.17--3.41 $\mu_B$\,f.u.$^{-1}$.

{Treating \lozfept{} FePt as a weak ferromagnet, MAE \textit{versus} FSM dependency could be mapped to the MAE \textit{versus} temperature dependency~\cite{haiMagneticPropertiesHard2003,edstromMagneticPropertiesFe1xCox2015}, employing the Callen and Callen model~\cite{callenPresentStatusTemperature1966}. 
Our analysis of the total Fe 3$d$ and Pt 5$d$ density of states at the majority spin channel pointed us towards a conclusion that treating \lozfept{} as a weak ferromagnet can be, at least, ambiguous -- and with the artificially decreased total moment situation gets even more complex. 
Mryasov \textit{et al.} have earlier shown that a precise classification of FePt could be not so simple~\cite{mryasovTemperaturedependentMagneticProperties2005}. 
Moreover, standard DFT as a zero kelvin model, omits a range of other temperature-dependent phenomena. 
Concluding, the mentioned procedure applied to other magnetic materials~\cite{edstromMagneticPropertiesFe1xCox2015} and experimental data regarding FePt~\cite{haiMagneticPropertiesHard2003} could not be reliably reproduced by the DFT method.}

Fig.~\ref{fig-fixed-spin} presents nearly identical MAE($m$) dependencies obtained near the equilibrium value with various exchange-correlation potentials, except for the exchange only approach.
However, exact values of the equilibrium total magnetic moment differ between the employed potentials.
This fact leads us to the conclusion, that observed wide distribution of the calculated MAE values for \lozfept{} in~\cite{wollochInfluenceAntisiteDefects2017} can be explained by the fact that different exchange-correlation potentials lead to different values of magnetic moment.
On the other hand, there is a system property independent of the choice of exchange-correlation potential, indicated by all considered exchange-correlation potential (except exchange only approach), which is an MAE maximum of about 20.3 MJ\,m$^{-3}$ observed for magnetic moment of about 2.9$\mu_B$.
In all cases, the equilibrium MAE results obtained for the optimized structure lie to the right of the MAE maximum. 
Hence, the modifications leading to the reduction of the \lozfept{} magnetization should increase the MAE to the hypothetical maximum value of 20.3~MJ\,m$^{-3}$ at 0~K.
A reduction in magnetization on the order of a few percent can be achieved, for example, by replacing a few percent of Fe with related elements exhibiting lower magnetic moment or completely non-magnetic, such as, for example, the 3$d$ elements Ni, Ti or V. 
Alternatively, one can try to slightly increase the Pt content at the expense of Fe or doping FePt with elements such as C, B and N, which should locate in the interstitial gaps.

It is worth noting that one would expect a monotonic increase in magnetic moments with decreasing temperature, leading to an MAE maximum at a few kelvins~\cite{haiMagneticPropertiesHard2003}.
Taking this fact into account, for room temperature applications, the goal should be to increase the magnetic moment of \lozfept{}, which should correspond to an increase of MAE.

The magnetostriction coefficient $\lambda_{001}$, determined from the Eq.~\ref{eq:magnetostriction}, is $\mathrm{4.3\times 10^{-4}}$. 
Magnetostriction coefficient $\lambda_S$ equals $3.4\times 10^{-5}$ -- $6.4\times 10^{-5}$ for disordered FePt thin films~\cite{spadaXrayDiffractionMossbauer2003,aboafMagneticTransportStructural1984}. Hence, the calculated value is an order of magnitude greater than that determined from the experiments, which is an expected behaviour, as the calculations conducted by us relate to a perfect infinite single crystal.
Compared to Terfenol-D (1.1-1.4$\times 10^{-3}$~\cite{sandlundMagnetostrictionElasticModuli1994}), the calculated magnetostriction coefficient of the \feptloz{} structure is an order of magnitude lower.

\section{Summary and Conclusions}

The magnetocrystalline anisotropy constants \textit{K$\mathrm{_1}$} and \textit{K$\mathrm{_2}$}, magnetic moments, and Curie temperatures were calculated for \lozfept{} using full-potential methods and, where possible, compared with equivalent ASA calculations. 
The results are in good agreement with the literature values of other calculations, though reproduce well-known discrepancy between DFT calculations and experiments. 
We have shown that the MAE results calculated with different exchange-correlation potentials correlate with the magnetic moment values, which explains the observed dispersion of the values determined so far.
The correlation of MAE($m$) FSM results near the equilibrium state is exceptional. 
For the magnetic moment reduced by about 10\%, we determined a theoretical maximum in MAE of 20.3~MJ\,m$^{-3}$, which is about 30\% higher than the theoretical equilibrium ground state value. 
We also derived the Curie temperature of \lozfept{} using multiple exchange-correlation functionals in the DLM approach and found it to be in decent agreement with the experimental value.
The evaluated magnetostriction coefficient $\lambda_{001}$ is in reasonable agreement with the measured values of disordered thin films of FePt alloy.

\section*{Acknowledgements}

We acknowledge the financial support of the National Science Centre Poland under the decision DEC-2018/30/E/ST3/00267.
The computations were in the main part performed on resources provided by the Poznan Supercomputing and Networking Center (PSNC).
We thank Paweł Leśniak and Daniel Depcik for compiling the scientific software and administration of the computing cluster at the Institute of Molecular Physics, Polish Academy of Sciences.
We thank J\'an Rusz for valuable discussion and suggestions.

\bibliography{fept_l10}

\end{sloppypar}
\end{document}